\documentstyle[11pt,newpasp_mod,twoside,epsf]{article}
\def\lsim{\mathrel{\rlap{\lower 4pt \hbox{\hskip 1pt $\sim$}}\raise 1pt \hbox
        {$<$}}}
\def\gsim{\mathrel{\rlap{\lower 4pt \hbox{\hskip 1pt $\sim$}}\raise 1pt \hbox
        {$>$}}}

\def\edcomment#1{\iffalse\marginpar{\raggedright\sl#1\/}\else\relax\fi}
\reversemarginpar

\begin{document}
\title{Hypernova Nucleosynthesis and Early Chemical Evolution}
 \author{Ken'ichi Nomoto, Keiichi Maeda, Hideyuki Umeda}
\affil{Department of Astronomy and Research Center for 
      the Early Universe, School of Science, University of Tokyo,
      Bunkyo-ku, Tokyo 113-0033, JAPAN}
%\author{}
%\affil{}

\begin{abstract}

We review the characteristics of nucleosynthesis in 'Hypernovae',
i.e., supernovae with very large explosion energies ($ \gsim 10^{52} $
ergs).  The hypernova yields compared to those of ordinary
core-collapse supernovae show the following characteristics: 1)
Complete Si-burning takes place in more extended region, so that the
mass ratio between the complete and incomplete Si burning regions is
generally larger in hypernovae than normal supernovae.  As a result,
higher energy explosions tend to produce larger [(Zn, Co)/Fe], smaller 
[(Mn, Cr)/Fe], and larger [Fe/O], which could explain the trend
observed in very metal-poor stars.  2) Si-burning takes place in lower
density regions, so that the effects of $\alpha$-rich freezeout is
enhanced.  Thus $^{44}$Ca, $^{48}$Ti, and $^{64}$Zn are produced more
abundantly than in normal supernovae.  The large [(Ti, Zn)/Fe] ratios
observed in very metal-poor stars strongly suggest a significant
contribution of hypernovae.  3) Oxygen burning also takes place in
more extended regions for the larger explosion energy.  Then a larger
amount of Si, S, Ar, and Ca ("Si") are synthesized, which makes the
"Si"/O ratio larger.  The abundance pattern of the starburst galaxy
M82 may be attributed to hypernova explosions.  Asphericity in the
explosions strengthens the nucleosynthesis properties of hypernovae
except for "Si"/O.  We thus suggest that hypernovae make important
contribution to the early Galactic (and cosmic) chemical evolution.

\end{abstract}

\section{Introduction}

Massive stars in the range of 8 to $\sim$ 100$M_\odot$ undergo
core-collapse at the end of their evolution and become Type II and
Ib/c supernovae (SNe II and SNe Ib/c).  Until recently, we have
considered supernovae with the explosion energies of $E =$ 1 - 1.5
$\times$ 10$^{51}$ ergs.  These energies have been best estimated from
SNe 1987A, 1993J, and 1994I, whose progenitors' masses have been
estimated to be 13 - 20 $M_\odot$ (e.g., Nomoto et al. 1993, 1994;
Blinnikov et al. 2000).

Recently, SN Ic 1998bw, which was associated with GRB980425 (Iwamoto
et al. 1998; Woosley et al. 1999), and SN Ic 1997ef (Iwamoto et
al. 2000; Mazzali et al. 2000) have been found to have such a large
kinetic explosion energy as $E \gsim 10^{52}$ erg.  This is more than
one order of magnitude larger than typical SNe, so that these objects
may be called "Hypernovae".  These SNe produced more $^{56}$Ni than
the average core collapse SN.  Their progenitors' masses are estimated
to be $M \gsim 25 M_\odot$.  These massive stars are likely to form
black holes, while less massive stars form neutron stars (see,
however, Wheeler et al. 2000).

We investigate the characteristics of nucleosynthesis in such
energetic core-collapse hypernovae, the systematic study of which has
not yet been done.  We examine both spherical and aspherical explosion
models and discuss their contributions to the Galactic chemical
evolution.

\section{Nucleosynthesis in Hypernova Explosions}

\subsection{Silicon and Oxygen Burning}

In core-collapse supernovae/hypernovae, stellar material undergoes
shock heating and subsequent explosive nucleosynthesis. Iron-peak
elements are produced in two distinct regions, which are characterized
by the peak temperature, $T_{\rm peak}$, of the shocked material.  For
$T_{\rm peak} > 5\times 10^9$K, material undergoes complete Si burning
whose products include Co, Zn, V, and some Cr after radioactive
decays.  For $4\times 10^9$K $<T_{\rm peak} < 5\times 10^9$K,
incomplete Si burning takes place and its after decay products include
Cr and Mn (e.g., Hashimoto et al. 1989; Woosley, \& Weaver 1995;
Thielemann et al. 1996).

We note the following characteristics of nucleosynthesis with very
large explosion energies (Nomoto et al. 2001):

1) Both complete and incomplete Si-burning regions shift outward in
mass compared with normal supernovae, so that the mass ratio between
the complete and incomplete Si-burning regions becomes larger.  As a
result, higher energy explosions tend to produce larger [(Zn, Co)/Fe],
smaller [(Mn, Cr)/Fe], and larger [Fe/O].  The elements synthesized in
this region such as $^{56}$Ni, $^{59}$Cu, $^{63}$Zn, and $^{64}$Ge
(which decay into $^{56}$Co, $^{59}$Co, $^{63}$Cu, and $^{64}$Zn,
respectively) are ejected more abundantly than in normal supernovae.  

2) In the complete Si-burning region of hypernovae, elements produced
by $\alpha$-rich freezeout are enhanced because nucleosynthesis
proceeds at lower densities (i.e., higher entropy) and thus a larger
amount of $^{4}$He is left.  Hence, elements synthesized through
capturing of $\alpha$-particles, such as $^{44}$Ti, $^{48}$Cr, and
$^{64}$Ge (decaying into $^{44}$Ca, $^{48}$Ti, and $^{64}$Zn,
respectively) are more abundant.

3) Oxygen burning takes place in more extended, lower density regions
for the larger explosion energy.  Therefore, more O, C, Al are burned
to produce a larger amount of burning products such as Si, S, and Ar.
Therefore, hypernova nucleosynthesis is characterized by large
abundance ratios of [Si/O], [S/O], [Ti/O], and [Ca/O].

\subsection {Aspherical Explosions}

Nakamura et al. (2001a) and Mazzali et al. (2001) have identified some
signatures of asymmetric explosion in the late light curve and spectra
of SN 1998bw.  Maeda et al. (2002) have examined the effect of
aspherical (jet-like) explosions on nucleosynthesis in hypernovae.
The progenitor model is the 16 $M_\odot$ He core of the 40 $M_\odot$
star and the explosion energy is $E$ = 1 $\times$ 10$^{52}$ ergs.

\begin{figure}
  \hspace*{0.1\textwidth}
  \begin{minipage}[t]{0.6\textwidth}
     \plotone{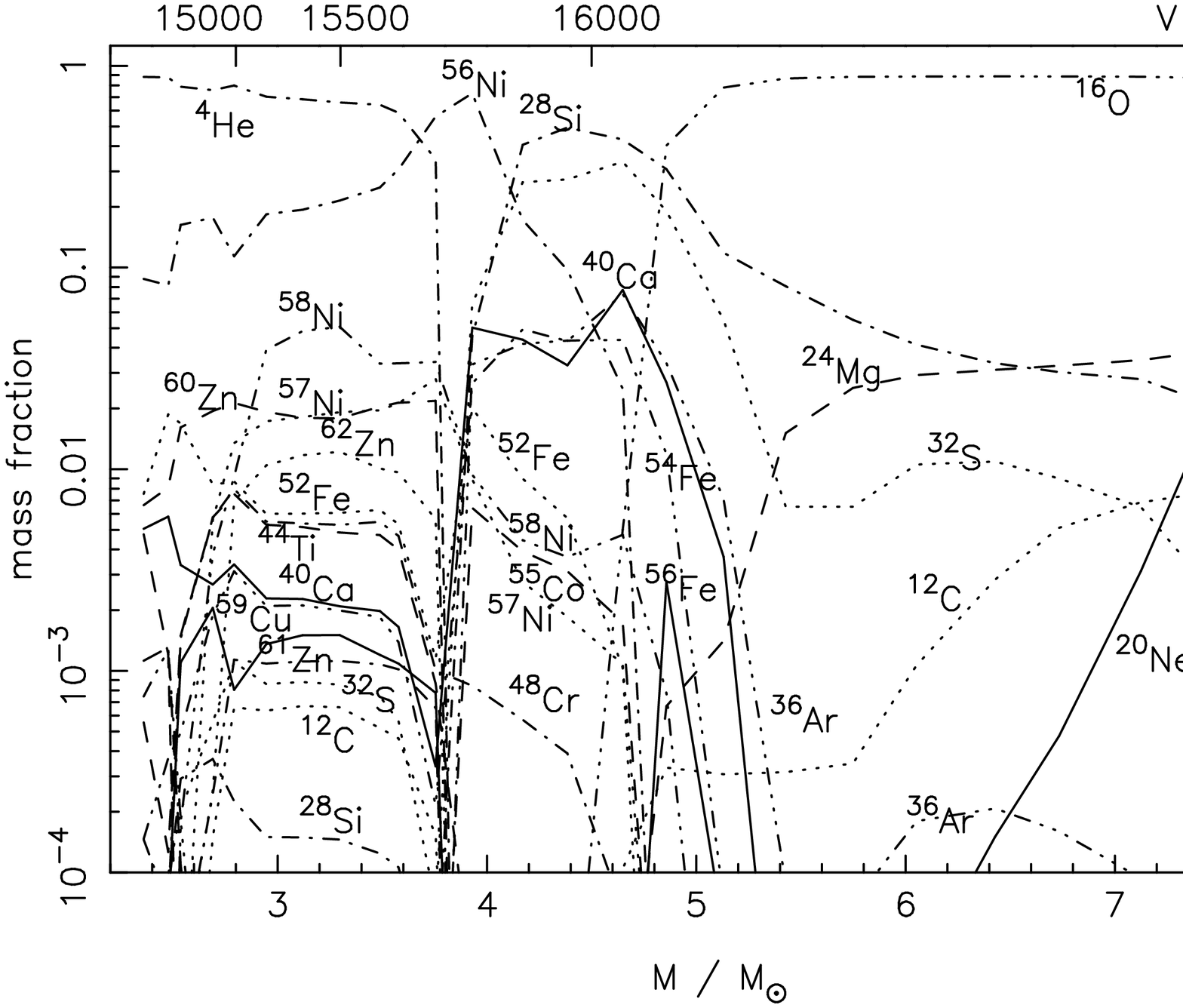}
  \end{minipage}\\
  \hspace*{0.1\textwidth}
  \begin{minipage}[t]{0.6\textwidth}
     \plotone{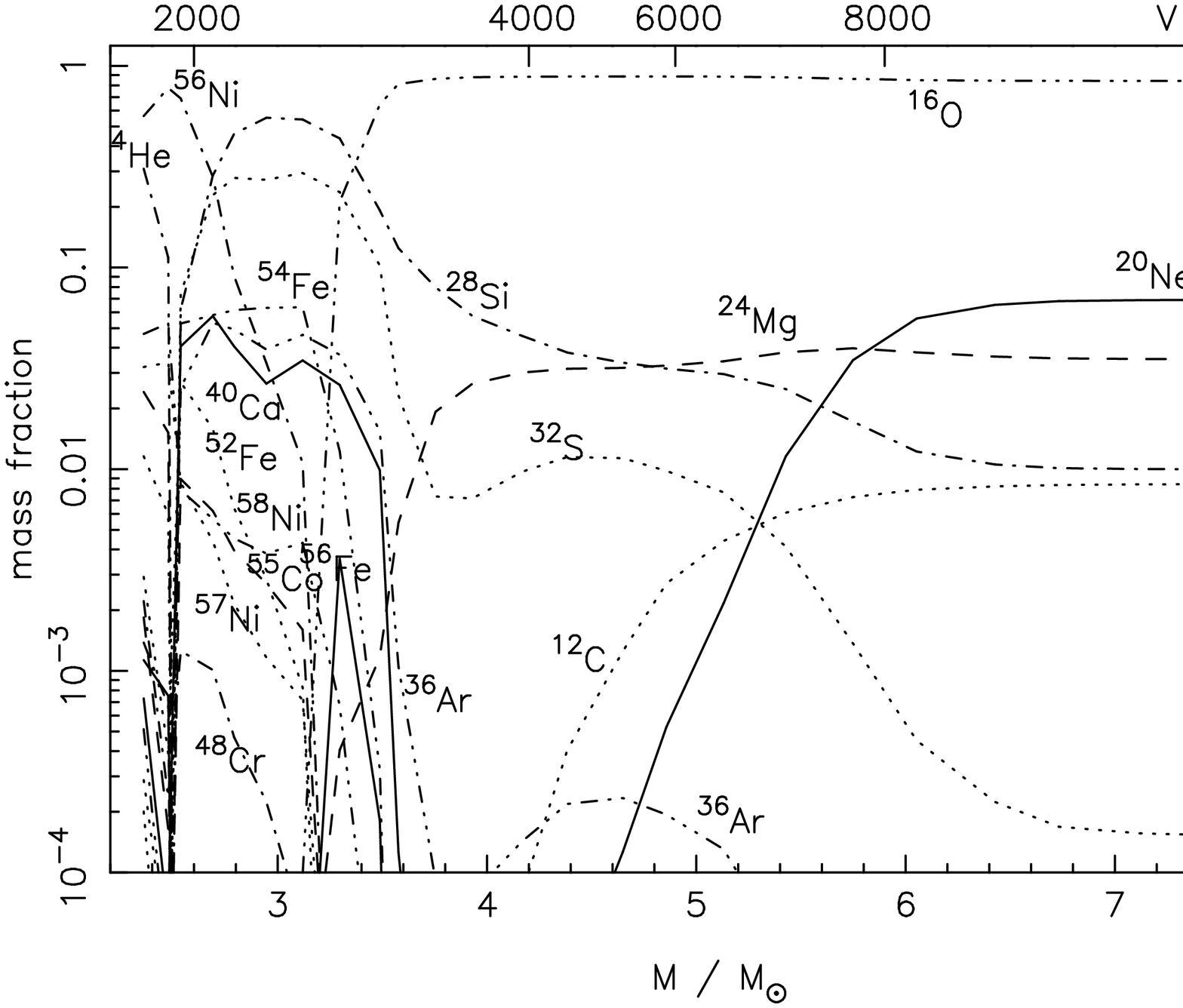}
  \end{minipage}
\caption{
The isotopic composition of the ejecta in the direction of 
the jet (upper panel) and perpendicular to it (lower panel).
The ordinate indicates the initial spherical Lagrangian coordinate ($M_r$) 
of the test particles (lower scale), and 
the final expansion velocities ($V$) of those particles (upper scale)
(Maeda et al. 2002).
\label{fig:nuc1d}}
\end{figure}

Figure 1 shows the isotopic composition of the ejecta of asymmetric
explosion model in the direction of the jet (upper panel) and
perpendicular to it (lower panel).  Figure 2 shows the 2D distribution
of $^{56}$Ni and $^{16}$O in the homologous expansion phase.

In the $z$-direction, where the ejecta carry more kinetic energy, the
shock is stronger and post-shock temperatures are higher.  Therefore,
larger amounts of $\alpha$-rich freeze-out elements, such as $^4$He,
$^{44}$Ti, and $^{56}$Ni are produced in the $z$-direction than in the
$r$-direction.

On the other hand, along the $r$-direction $^{56}$Ni is produced only
in the deepest layers, and the elements ejected in this direction are
mostly the products of hydrostatic nuclear burning stages (O) with
some explosive oxygen-burning products (Si, S, etc). 

In the spherical case, Zn is produced only in the deepest layer, while
in the aspherical model, the complete silicon burning region is
elongated to the $z$ (jet) direction, so that [Zn/Fe] is enhanced
irrespective of the mass cut.  On the other hand, $^{55}$Mn, which is
produced by incomplete silicon burning, surrounds $^{56}$Fe and
located preferentially in the $r$-direction.

In this way, larger asphericity in the explosion leads to larger
[Zn/Fe] and [Co/Fe], but to smaller [Mn/Fe] and [Cr/Fe].  Then, if the
degree of the asphericity tends to be larger for lower [Fe/H], the
trends of [Zn, Co, Mn, Cr/Fe] follow the ones observed in metal-poor stars, 
as discussed later. 

\begin{figure}
  \hspace*{0.1\textwidth}
  \begin{minipage}[t]{0.4\textwidth}
    \plotone{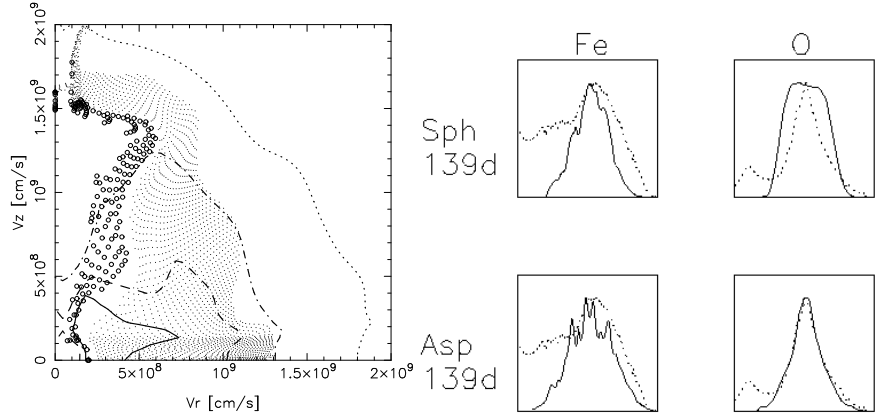}
  \end{minipage}
  \begin{minipage}[t]{0.45\textwidth}
    \plotone{line_S+A2.epsi}
  \end{minipage}
\caption{
Left: The distribution of $^{56}$Ni (open circles) and $^{16}$O (dots).
The open circles and 
the dots denote test particles in which the mass fraction 
of $^{56}$Ni and $^{16}$O, respectively, exceeds 0.1. 
The lines are density contours at the level of 0.5 (solid), 0.3 (dashed),
0.1 (dash-dotted), and 0.01 (dotted) of the max density, respectively
(Maeda et al. 2002).
Right: The profiles of the Fe-blend (left panels) and of  
O {\small I}] 6300, 6363 \AA\ (right panels) 
viewed at 15$^{\circ}$ from the jet direction (Maeda et al. 2002).  
The top panels show the profiles of the spherically symmetric model.
The observed lines at a SN rest-frame epoch of 139 days are also plotted for 
comparison (dotted lines, Patat et al. 2001).
\label{fig:nuc2d}}
\end{figure}

\subsection{Nebula Spectra of SN 1998bw}

In order to verify the observable consequences of an axisymmetric
explosion, Maeda et al. (2002) calculated the profiles of the
Fe-dominated blend near 5200\AA, and of O {\small I}] 6300, 6363\AA.
These are the lines that deviate most from the expectations from a
spherically symmetric explosion (Mazzali et al. 2001).  

The iron and oxygen profiles viewed at an angle of 15$^{\circ}$ from
the jet direction are found to be most consistent with the observed
spectrum on day 139 in Figure 2.  When the degree of asphericity is
high and the viewing angle is close to the jet direction, the
component iron lines in the blend have double-peaked profiles, the
blue- and red-shifted peaks corresponding to Fe-dominated matter
moving towards and away from us, respectively.  Because of the high
velocity of Fe, the peaks are widely separated, making the blend wide. 
This is the case for the synthetic Fe-blend shown in Figure 2.  In
contrast, the oxygen line is narrower and has a sharper peak, because
O is produced mostly in the $r$-direction, at lower velocities and
with a less aspherical distribution.

\section{Signatures of Hypernova Nucleosynthesis in Chemical Evolution}

The abundance pattern of metal-poor stars with [Fe/H] $< -2$ provides
us with very important information on the formation, evolution, and
explosions of massive stars in the early evolution of the galaxy.

In the early galactic epoch when the galaxy is not yet chemically
well-mixed, [Fe/H] may well be determined by mostly a single SN event 
(Audouze \& Silk 1995). The formation of metal-poor stars is supposed
to be driven by a supernova shock, so that [Fe/H] is determined by the
ejected Fe mass and the amount of circumstellar hydrogen swept-up by
the shock wave (Ryan, Norris, \& Beers 1996).  
Then, hypernovae with larger $E$
are likely to induce the formation of stars with smaller [Fe/H],
because the mass of interstellar hydrogen swept up by a hypernova is
roughly proportional to $E$ (Ryan et al. 1996; Shigeyama \& Tsujimoto 1998) 
and the ratio of the ejected iron mass to
$E$ is smaller for hypernovae than for canonical supernovae.

\subsection {Zn, Co, Mn, Cr}

The observed abundances of metal-poor halo stars show quite
interesting pattern.  There are significant differences between the
abundance patterns in the iron-peak elements below and above [Fe/H]$
\sim -2.5$ - $-3$.  

1) For [Fe/H]$\lsim -2.5$, the mean values of [Cr/Fe] and [Mn/Fe]
decrease toward smaller metallicity, while [Co/Fe] increases 
(McWilliam et al. 1995; Ryan et al. 1996). 

2) [Zn/Fe]$ \sim 0$ for [Fe/H] $\simeq -3$ to $0$ (Sneden, Gratton, \&
Crocker 1991), while at [Fe/H] $< -3.3$, [Zn/Fe] increases toward
smaller metallicity (Figure 3; Primas et al. 2000; Blake et al. 2001).

These trends cannot be explained with the conventional chemical
evolution model that uses previous nucleosynthesis yields.

The larger [(Zn, Co)/Fe] and smaller [(Mn, Cr)/Fe] in the supernova
ejecta can be realized if the mass ratio between the complete Si
burning region and the incomplete Si burning region is larger, or
equivalently if deep material from complete Si-burning region is
ejected by mixing or aspherical effects.  This
can be realized if (1) the mass cut between the ejecta and the
collapsed star is located at smaller $M_r$ (Nakamura et al. 1999), (2)
$E$ is larger to move the outer edge of the complete Si burning region
to larger $M_r$ (Nakamura et al. 2001b), or (3) asphericity in the
explosion is larger.

Also a large explosion energy $E$ results in the enhancement of the
local mass fractions of Zn and Co, while Cr and Mn are not enhanced 
(Umeda \& Nomoto 2001).  Therefore, if hypernovae made significant
contributions to the early Galactic chemical evolution, it could
explain the large Zn and Co abundances and the small Mn and Cr
abundances observed in very metal-poor stars.

\vspace{0.5cm}

\begin{figure}
%  \begin{minipage}[t]{0.5\textwidth}
%    \includegraphics[height=.3\textheight]{znfe2}
%  \end{minipage}
%  \begin{minipage}[t]{0.5\textwidth}
%    \includegraphics[height=.3\textheight]{znfeMEXP}
%  \end{minipage}
  \plottwo{znfe2.epsi}{znfeMEXP.epsi}
\caption{
Left: Observed abundance ratios of [Zn/Fe].  These data are taken
from Primas et al. (2000) (filled circles), Blake et al. (2001) 
(filled square) and from Sneden et al. (1991) (others) 
(Umeda \& Nomoto 2001).
\newline
Right: The maximum [Zn/Fe] ratios as a function of $M$ and $E_{51} =
E/10^{51}$ergs (Umeda \& Nomoto 2001).  The observed large [Zn/Fe]
ratio in very low-metal stars ([Fe/H] $<-2.6$) found in
Primas et al. (2000) and Blake et al. (2001) are represented 
by a thick arrow.
\label{fig:znfe}}
\end{figure}

The dependence of [Zn/Fe] on $M$ and $E$ is summarized in Figure 3.  
Models with $E_{51} = E/10^{51}$ergs do not 
produce sufficiently large [Zn/Fe].  To be compatible with the
observations of [Zn/Fe] $\sim 0.5$, the explosion energy must be much
larger, i.e., $E_{51} \gsim 20$ for $M \gsim 20 M_\odot$, i.e.,
hypernova-like explosions of massive stars ($M \gsim 25 M_\odot$) with
$E_{51} > 10$ are responsible for the production of Zn.

\subsection{Ni}

 The observed trend of another iron-peak element, Ni, is also
interesting. Unlike the elements we have focused on, [Ni/Fe] of
metal-poor stars shows no clear trend (see e.g., Ryan et al. 1996;
Nakamura et al. 1999).  Theoretically, this is understood as the fact
that Ni is produced abundantly by both complete and incomplete
Si-burning.  Recently, Elliison et al. (2001) observed DLA abundance
and found [Co/Fe] $>0$, which is similar to the metal-poor halo stars.
They, on the other hand, did not find oversolar [Ni/Fe].  Similar
results have also been found by Norris et al. (2001), who observed
abundances of five halo stars with [Fe/H] $\lsim -3.5$.  They
discussed that the results are inconsistent with the predictions of
Nakamura et al. (1999), where the enhancement of [Co/Fe] appears to be
accompanied by the enhancement of [Ni/Fe].

 We note that the increase in [Ni/Fe] along with the increase in
[Co/Fe] is not significant in the hypernova models by Umeda \& Nomoto
(2001); for example, Co/Fe in the hypernova is larger by a factor of
17 than a normal SN II, while Ni is larger only by a factor of 1.8.

 The apparent difference from the results in Nakamura et al. (1999)
can be understood as follows. In Nakamura et al. (1999), the explosion
energy was fixed to be $E_{51}=1$. They obtained the larger Co/Fe
ratio for more massive SNe II by assuming ``deeper'' mass-cuts so that
$Y_e$ in the explosive burning region is smaller $(Y_e \simeq 0.495$).
For smaller $Y_e$, the Co abundance is larger, but the abundance of Ni
(especially $^{58}$Ni) is enhanced by a larger factor than Co.
Therefore, the increase in [Ni/Fe] with [Co/Fe] was unavoidable,
unless neutrinos substantially enhance $Y_e$ in the deep complete Si
burning region.

In Umeda \& Nomoto models, mass-cuts of the larger Co (and Zn) models
are not deeper in $M_r$, and $Y_e$ in the complete Si burning region is
not small.  This is because they assume larger explosion energies for
more massive stars, which shifts the mass-cut outwards in $M_r$.  As a
result, the dominant Ni isotope is $^{60}$Ni. In our model, with
increasing $E$, the abundances of Co and Zn increase more
than $^{60}$Ni. Therefore, Co and Zn abundances can be enhanced
without appreciable increase in the Ni abundance.  In this sense, the
abundance trends of very metal-poor stars is better explained with
hypernova models rather than the simple ``deep'' mass-cut models
(Nakamura et al. 1999).

\subsection{Pair Instability Supernovae ?}

     One may wonder whether the abundance anomaly of iron-peak
elements may be related to the peculiar IMF of Pop III stars.  It is
quite likely that the IMF of Pop III stars is different from that of
Pop I and II stars, and that more massive stars are abundant for Pop
III. Nakamura \& Umemura (1999) discussed that the
IMF of Pop III and very low metal stars may have a peak at even larger
masses, around $\sim (1 $- few)$\times 100M_\odot$.  If $M\lsim
130M_\odot$, then these stars are likely to form black holes either
without explosion or with energetic explosions.  The nucleosynthesis
of the latter case may not be so different from the models considered
here.  This might favor the scenario that invokes the hypernova-like
explosions for large [Zn/Fe].

     If stars are even more massive than $\sim 150 M_\odot$, these
stars become pair-instability SNe (PISNe) and their nucleosynthesis is
different from core-collapse SNe.  In particular, PISNe produce [Zn/Fe]
$< -1.5$, because in PISNe, iron peak elements are mostly produced by
incomplete Si burning so that the mass fraction of complete Si burning
elements is much smaller than SNe II (Umeda \& Nomoto 2001).  We thus
conclude that PISNe are unlikely to produce a large enough Zn/Fe ratio
to explain the observations.

\section {Starburst Galaxy M82 and Hypernovae}

X-ray emissions from the starburst galaxy M82 were observed with ASCA
and the abundances of several heavy elements were obtained (Tsuru et
al. 1997).  Tsuru et al. (1997) found that the overall metallicity of
M82 is quite low, i.e., O/H and Fe/H are only 0.06 - 0.05 times solar,
while Si/H and S/H are $\sim$ 0.40 - 0.47 times solar.  This implies
that the abundance ratios are peculiar, i.e., the ratio O/Fe is about
solar, while the ratios of Si and S relative to O and Fe are as high
as $\sim$ 6 - 8.  These ratios are very different from those ratios in
SNe II.  Compared with normal SNe II, the important characteristic of
hypernova nucleosynthesis is the large Si/O, S/O, and Fe/O ratios.
Figure 4 shows the good agreement between the hypernova model
($E_{51}=$ 30) and the observed abundances in M82.

Hypernovae could also produce larger $E$ per oxygen mass than normal
SNe II, as required for M82.  We therefore suggest that hypernova
explosions may make important contributions to the metal enrichment
and energy input to the interstellar matter in M82.  The age of
starburst activity is estimated to be $\lsim 10^7$ years (Stickland
2001), which is so young that only massive stars ($M >$ 25
$M_\odot$) contributed to nucleosynthesis in M82.

\begin{figure}
 \hspace*{0.2\textwidth}
  \begin{minipage}[t]{0.6\textwidth}
     \plotone{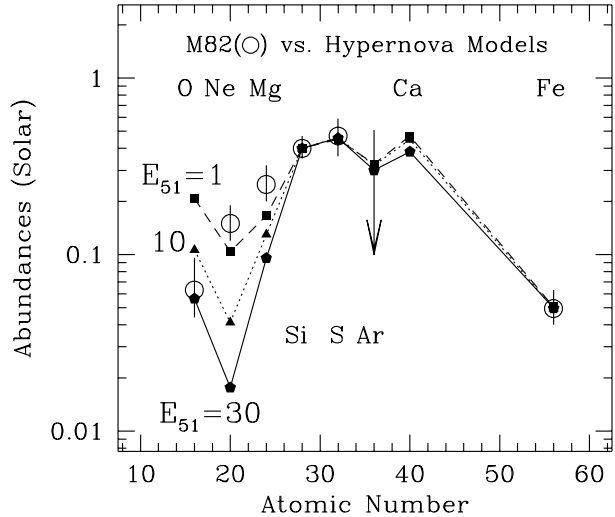} 
%     \plotone{25Z0abun1CL.epsi} 
% use 25Z0abun1BW.epsi if you want to make it black-white. 
  \end{minipage}
\caption{
Abundance patterns in the ejecta of the 25$M_\odot$ metal-free SN II and
hypernova models compared with abundances (relative to the solar
values) of M82 observed with ASCA (open circles: Tsuru et al. 1997).
The filled square, triangle, and pentagons represent $E_{51}$=1, 10,
and 30 models, respectively.
\label{fig:m82}}
\end{figure}

\section{Hypernovae vs. Type Ia Supernovae}

The large [Fe/O] observed in some metal-poor stars and galaxies might
possibly be the indication of Hypernovae rather than Type Ia
supernovae (SNe Ia).

To distinguish between Hypernova (Nakamura et al. 2001b; Umeda \&
Nomoto) and SN Ia (Iwamoto et al. 1999; Nomoto et al. 1997b)
nucleosynthesis, following abundance ratios are useful:
\bigskip\bigskip

[Ti/Fe] $\sim$ 0.3 - 0.5 in Hypernovae; $<$ 0 in SNe Ia.

[Mn/Fe] $<$ 0 in Hypernovae; $>$ 0 in SNe Ia.

[Ni/Fe] $\sim$ 0 - 0.3 in Hypernovae; $\sim$ 0 - 0.5 in SNe Ia.

[Zn/Fe] $\sim$ 0 - 0.5 in Hypernovae; $<$ 0 in SNe Ia.

[Si/Fe] $\sim$ 0.3 - 0.5 in Hypernovae; $\sim -0.5$ - 0 in SNe Ia.

\bigskip\noindent
To reproduce the solar abundance ratios with the combination of SN II
and SN I products, there exist certain constraints on the model
abundances of SNe Ia.  For example, [Si/Fe] $\sim$ 0.3 in SNe II
(Nomoto et al. 1997a) so that [Si/Fe] $< -0.2$ in SNe Ia (Iwamoto et
al. 1999).  For a system with a non-solar abundance pattern, this
constraint would be weaker.

\section{Concluding Remarks}

We have shown that signatures of hypernova nucleosynthesis are seen in
the large [(Ti, Zn)/Fe] ratios in very metal poor stars and the large
Si/O and Fe/O ratios in the starburst galaxy M82.  (See also the
abundance pattern in X-ray Nova Sco; Israelian et al. 1999;
Podsiadlowski et al. 2001).  These properties of hypernova
nucleosynthesis suggest that hypernovae of massive stars may make
important contributions to the Galactic (and cosmic) chemical
evolution, especially in the early low metallicity phase.  This may be
consistent with the suggestion that the IMF of Pop III stars is
different from that of Pop I and II stars, and that more massive stars
are abundant for Pop III.

\begin{acknowledgments}

This work has been supported in part by the grant-in-Aid for
Scientific Research (07CE2002, 12640233) of the Ministry of Education,
Science, Culture, and Sports in Japan.

\end{acknowledgments}

\end{document}